\documentclass{article}
\usepackage[T1]{fontenc}
\usepackage[ansinew]{inputenc}
\usepackage{graphicx}
\title{Hyperbolic subdiffusive impedance}

\author{Tadeusz Koszto{\l}owicz$^1$ and Katarzyna D. Lewandowska$^2$}

\date{\footnotesize{$^1$Institute of Physics, University of Kielce,\\
         ul. \'Swi\c{e}tokrzyska 15, 25-406 Kielce, Poland, \\
         e-mail: tkoszt@pu.kielce.pl\\
        $^2$Department of Physics and Biophysics, Medical University of
         Gda\'nsk,\\ ul. D\c{e}binki 1, 80-211 Gda\'nsk, Poland, \\
         e-mail: kale@amg.gda.pl}}

\begin{document}
\maketitle

\begin{abstract}
We use the hyperbolic subdiffusion equation with fractional time derivatives (the generalized Cattaneo equation) to study the transport process of electrolytes in media where subdiffusion occurs. In this model the flux is delayed in a non-zero time with respect to the concentration gradient. In particular, we obtain the formula of electrochemical subdiffusive impedance of a spatially limited sample in the limit of large and of small pulsation of the electric field. The boundary condition at the external wall of the sample are taken in the general form as a linear combination of subdiffusive flux and concentration of the transported particles. We also discuss the influence of the equation parameters (the subdiffusion parameter and the delay time) on the Nyquist impedance plots.
\end{abstract}

\section{\label{i}Introduction}

Subdiffusion occurs in systems where mobility of particles is
significantly hindered due to internal structure of the medium, as in porous media, gels or amorphous semiconductors \cite{mk,kdm}. The subdiffusion is characterized by a time dependence of the mean square displacement of transported particle $\left\langle \Delta x^2\right\rangle=\frac{2D_\alpha t^{\alpha}}{\Gamma\left(1+\alpha\right)}$, where $D_\alpha$ is the subdiffusion coefficient measured in the units $m^2/s^{\alpha}$ and $\alpha$ is the subdiffusion parameter of a value within $0<\alpha<1$ range. For $\alpha=1$ one deals with the normal diffusion. 

The subdiffusion has been recently extensively studied. While the phenomenon is theoretically rather well understood, there are very few reported experimental investigations (e.g. \cite{kdm}).  
The method of impedance spectroscopy was used to experimentally study subdiffusion in porous media such as nanopore electrode \cite{b}, cement \cite{dnp,c1,c2}, tooth enamel \cite{rg} and gels \cite{jpl}. 
The theoretical analysis of subdiffusion impedance was presented by \cite{bc} who used the following  parabolic subdiffusion equation (PSE) with fractional time derivative. The equation reads as
	\begin{equation}\label{pse}
\frac{\partial C(x,t)}{\partial t}=D_\alpha\frac{\partial^{1-\alpha}}{\partial t^{1-\alpha}}\frac{\partial^2C(x,t)}{\partial x^2} ,
	\end{equation}
where the Riemann-Liouville fractional time derivative, which is defined for $\alpha>0$ as \cite{p,os}
    \begin{displaymath}
\frac{\partial^{\alpha}f(t)}{\partial
t^{\alpha}}=\frac{1}{\Gamma(n-\alpha)}\frac{\partial^{n}}{\partial
t^{n}}\int_{0}^{t}dt'\frac{f(t')}{(t-t')^{1+\alpha-n}} .
    \end{displaymath}
For $\alpha=1$ Eq. (\ref{pse}) converts into the normal diffusion equation.

For the initial condition $C(x,0)=\delta(x)$, where $\delta$ is the Dirac-delta function, the solution of Eq. (\ref{pse}) (the Green's function) has non-zero values for any $x$ and $t$ ($t>0$). Thus, even for small times, a finite amount of the substance exists at very large distances from the origin, what can be interpreted as infinite propagation velocity of some of the diffusing particles. To avoid this 'unphysical property' Cattaneo derived the hyperbolic normal diffusion equation which Green's function has non-zero values for finite $x$ \cite{cattaneo,cm}. The equation is based on the assumption, that the flux is delayd by time period $\tau$ with respect to the concentration gradient.
For many 'typical systems' (as the membrane one) it is hard to observe the difference between the solutions of parabolic and hyperbolic (sub)diffusion equations even for relatively large values of $\tau$ \cite{hs}. However, in some processes the non-zero parameter $\tau$ plays crucial role. The example is the diffusion in a system where boundary conditions are given by functions quickly changing in time. Such a situation occurs in the electrochemical system with (sub)diffusion impedance. As far as we know, the Cattaneo equation was used to study electrochemical impedance only for a system where normal diffusion occurs \cite{criado,rbgmcc}, except our work \cite{lk} where the subdiffusion impedance was considered in a system with fully absorbing wall. Articles published so far mostly concentrated on homogeneous systems, however, a more complex system containing few diffusion layers was also studied in \cite{freger}.   

In this paper we present a theoretical foundation for studies of subdiffusion impedance using a hyperbolic equation. We apply the hyperbolic Cattaneo equation with the fractional time derivative to model the subdiffusion impedance of a homogeneous sample of finite thickness, where the boundary condition at the sample surface is assumed as linear combination of flux and concentration. We find an influence of the parameters $\alpha$ and $\tau$ on the final formula describing the impedance of the subdiffusive medium, particularly for high and for low ac-voltage frequency. 

\section{\label{gce}The generalized Cattaneo equation}

The phenomenological derivation of Cattaneo equation is based on the assumption that the flux of the particles $J$ is not generated by the concentration gradient instantaneously (as in the process described by parabolic diffusion equation), but it is delayed in time by $\tau$ \cite{cattaneo}, what is reflected by the following relation
	\begin{equation}\label{ndyfs}
J(x,t+\tau)=-D_{\alpha}\frac{\partial C(x,t)}{\partial x} .
	\end{equation}
Expanding the left-hand side of Eq.~(\ref{ndyfs}) into power series with respect to $\tau$ and assuming that the parameter $\tau$ is sufficiently small, one gets
	\begin{equation}\label{nszers}
J(x,t)+\tau\frac{\partial J(x,t)}{\partial t}=-D_{\alpha}\frac{\partial C(x,t)}{\partial x} .
	\end{equation}
Combining Eq.~(\ref{nszers}) with the continuity equation
    \begin{equation}\label{ce}
\frac{\partial C(x,t)}{\partial t}=-\frac{\partial
J(x,t)}{\partial x} ,
    \end{equation}
one obtains the hyperbolic normal diffusion equation
	\begin{equation}\label{rowCatt}
\tau\frac{\partial^2C(x,t)}{\partial t^2}+\frac{\partial C(x,t)}{\partial t}=D_\alpha\frac{\partial^2C(x,t)}{\partial x^2} .
	\end{equation}
The parabolic subdiffusion equation can be derived form Continuous Time Random Walk formalism or using the phenomenological approach. In the letter case one sets the Riemann-Liouville fractional time derivative of the order $1-\alpha$ in the right-hand side of Eq. (\ref{ndyfs}) (with $\tau=0$) or replaces the time derivative of the first order in Eq. (\ref{ce}) to Caputo fractional time derivative of the order $\alpha$. In similar way one can obtain the hyperbolic subdiffusion equation. In the paper \cite{cm} there were proposed three possible generalizations of the Cattaneo equation (\ref{rowCatt}) to the one with the fractional time derivative, each one supported by a different scheme. In each of them the fractional time derivative replaces the derivative of natural order in Eqs (\ref{nszers}) or (\ref{ce}), or it is added to the right-hand side of Eq. (\ref{ndyfs}). The schemes provide a different forms of hyperbolic subdiffusion equation, which are not equivalent of each other. In the next section we use the last scheme - which is the most natural in our opinion - to derive the hyperbolic subdiffusion equation.
Thus, the generalized Fick equation is given by
	\begin{equation}\label{dyfs}
J(x,t+\tau)=-D_{\alpha}\frac{\partial^{1-\alpha}}{\partial t^{1-\alpha}}\frac{\partial C(x,t)}{\partial x} .
	\end{equation}
Equation (\ref{dyfs}) ensures that changes of the flux due to the concentration gradient are delayed by the time $\tau$. 
Assuming that $\tau\ll t$ and keeping linear terms with respect to $\tau$ in the series expansion of l.h.s. of Eq.~(\ref{dyfs}), we get
\begin{equation}\label{szers}
J(x,t)+\tau\frac{\partial J(x,t)}{\partial t}=-D_{\alpha}\frac{\partial^{1-\alpha}}{\partial t^{1-\alpha}}\frac{\partial C(x,t)}{\partial x} .
	\end{equation}
Combining Eq.~(\ref{szers}) with the continuity equation (\ref{ce}),
we obtain the generalized Cattaneo equation
	\begin{equation}\label{srowCatt}
\tau\frac{\partial^2C(x,t)}{\partial t^2}+\frac{\partial C(x,t)}{\partial t}=D_\alpha\frac{\partial^{1-\alpha}}{\partial t^{1-\alpha}}\frac{\partial^2C(x,t)}{\partial x^2} .
	\end{equation}

\section{\label{s}The system}

\begin{figure}[h!]
	\centering
		\includegraphics[width=10cm]{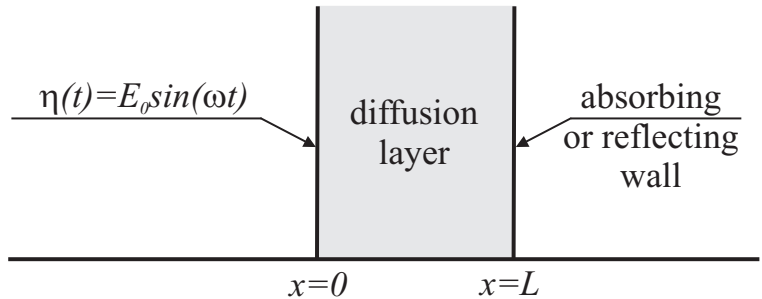}
	\caption{The system under considerations, $\eta$ denotes the overvoltage, $E$ - its amplitude.}
	\label{fig:uklad}
\end{figure}

Let us assume that at $x=0$ there is 
the oscillating overvoltage $\eta(t)=E\sin(\omega t)$ which causes the oscillation of the concentration on the surface layer according to the formula	
	\begin{equation}\label{oscst}
\left.\eta\right|_{x=0}(t)=\left(\frac{d\eta}{dC}\right)_{eq}C(0,t) ,
	\end{equation}
where $eq$ denotes a derivative computed in the local equilibrium. 
Thus, we have 
	\begin{equation}\label{wbdyfn}
C(0,t)=C_0\sin(\omega t) ,
	\end{equation}
where $C_0=R_WqAE$, $R_{W}=\frac{1}{qA}\left(\frac{d\eta}{dC}\right)_{eq}$.
The conduction current $I(t)$ at $x=0$ corresponds to the flux of diffusing particles $J(0,t)$
	\begin{equation}\label{nat}
I(t)=qAJ(0,t) ,	
	\end{equation}
where $q$ is the charge of diffusing particle and $A$ is the area of the sample surface.
The surface located at $x=L$ can be treated as fully absorbing, partially absorbing or fully reflecting wall (Fig.\ref{fig:uklad}). 

The second boundary condition, fixed at the wall $x=L$, can be chosen in different ways depending on the properties of the wall. Usually, for the system described by parabolic (sub)diffusion equation one can adopt the boundary conditions as follows:

\begin{itemize}
	\item
 For the fully absorbing wall one gets \cite{chan}
	\begin{equation}\label{bcfaw}
C(L,t)=0 .
	\end{equation}
	\item
For the fully reflecting wall there is \cite{chan}
	\begin{equation}\label{bcfrw}
J(L,t)=0 .
	\end{equation}
	\item
For the partially absorbing wall the particle absorbed by the wall cannot return to the system and the boundary condition is given by \cite{rn,koszt2001a}
	\begin{equation}\label{bcpaw}
J(L,t)=\left.\kappa\frac{\partial C(x,t)}{\partial x}\right|_{x=L} .
	\end{equation}
\end{itemize}
The boundary condition for the fully absorbing or fully reflecting wall was studied in majority works  \cite{mcdonald,bc}, however, the radiation boundary condition (\ref{bcpaw}) was also used \cite{criado,bisquert1999,bisquert1998}.
In this work we assume that the boundary condition at $x=L$ is given in a general form and it is a linear combination of the flux and concentration, where the flux is delayed in time by $\tau$ with respect to the concentration. 
So, the boundary condition is
	\begin{equation}\label{bc}
a_L J(L,t+\tau)+b_L C(L,t)=0 .
	\end{equation}

\section{\label{di}Diffusion impedance}

The impedance of electrochemical system $Z(i\omega)$ can be defined as its response to a voltage or current perturbation from a steady-state situation \cite{mcdonald,jacobsen} 
	\begin{equation}\label{dimp}
Z(i\omega)=\frac{\hat{\eta}(i\omega)}{\hat{I}(i\omega)} ,
	\end{equation}
where $\hat{\eta}(i\omega)$ and $\hat{I}(i\omega)$ are the Laplace transforms of the overvoltage and current perturbation. $\omega$  is the angular frequency $\omega=2\pi f$, where $f$ is frequency. 
A plot of the real and imaginary parts of the impedance in the complex plane $(ReZ,-ImZ)$ as the frequency (treated as a parameter) is swept over a given range is called the Nyquist plot. 

The impedance of the diffusion layer is called the Warburg impedance. For the layer of the infinite thickness the impedance is defined as
	\begin{equation}\label{impW}
Z(i\omega)=\frac{R}{\sqrt{i\omega}}=\frac{R}{\sqrt{2\omega}}(1-i) ,
	\end{equation}
where $R$ is the diffusion resistance. On the Nyquist plot the Warburg impedance is presented by the straight half line with the slop angle $\pi/4$ passing through the origin of coordinates. In real systems the diffusion layer has a finite thickness. Let us assume, that the diffusion layer is bordered by planes localized at $x=0$ and $x=L$. The perturbation of the voltage is applied to the medium at $x=0$. The characteristic angular frequency is defined as
	\begin{equation}\label{pulukl}
\omega_{d}\equiv\frac{D}{L^{2}} ,
	\end{equation}
where $D$ is the diffusion coefficient. The frequency (\ref{pulukl}) is proportional to the inverse of the average time necessary for an ion to cross the sample thickness. When $\omega\gg\omega_{d}$ the size of the sample plays no role in the ion diffusion and the impedance is the Warburg impedance (\ref{impW}). However, for a low frequency the ions can be absorbed by the opposite wall before they change direction of their movement.

 From Eqs.~(\ref{dimp}), (\ref{oscst}) and (\ref{nat}) one obtains the relation
	\begin{equation}\label{impststr}
Z(i\omega)=R_{W}\frac{\hat{C}(0,i\omega)}{\hat{J}(0,i\omega)} .
	\end{equation}

\section{\label{si}Subdiffusion impedance}

We assume that the transport process is described by the Cattaneo equation
(\ref{rowCatt}) with the following initial conditions
	\begin{equation}\label{wp}
C(x,0)=\left.\frac{\partial C(x,t)}{\partial t}\right|_{t=0}=0.
	\end{equation}
The Laplace transform of (\ref{rowCatt}) for the initial conditions (\ref{wp}) is
	\begin{equation}\label{tLrC}
\tau s^{2}\hat{C}(x,s)+s\hat{C}(x,s) =D_{\alpha}s^{1-\alpha}\frac{d^{2}\hat{C}(x,s)}{dx^{2}}.
	\end{equation}
The general solution of Eq.~(\ref{tLrC}) is
	\begin{equation}\label{rozwrC}
\hat{C}(x,s)=B_1\exp\left[\gamma(s)x\right]+B_2\exp\left[-\gamma(s)x\right] , 
	\end{equation}
where
	\begin{equation}\label{gamma}
\gamma(s)=\frac{s^{\alpha/2}}{\sqrt{D_{\alpha}}}\sqrt{1+\tau s} .
	\end{equation}
The Laplace transform of the flux is
	\begin{equation}\label{tLs}
\hat{J}(x,s)=-D_{\alpha}\frac{s^{1-\alpha}}{1+\tau s}\frac{d\hat{C}(x,s)}{dx} .
	\end{equation}
Combinig Eqs.~(\ref{impststr}), (\ref{bc}) and (\ref{rozwrC})--(\ref{tLs}) we obtain
	\begin{equation}\label{imp2}
Z(s)=R_W\frac{1}{\lambda(s)}\left[\frac{b_L \sinh(\gamma(s)L)-a_L \lambda(s)\cosh(\gamma(s)L)}{b_L \cosh(\gamma(s)L)-a_L \lambda(s)\sinh(\gamma(s)L)}\right] ,
	\end{equation}
where
	\begin{equation}\label{lambda}
\lambda(s)=s^{1-\alpha/2}\sqrt{\frac{D_{\alpha}}{1+\tau s}} .
	\end{equation}
For the layer with infinite thickness $(L\rightarrow\infty)$ the impedance (\ref{imp2}) has the following  form
	\begin{equation}\label{imp3}
Z(i\omega)=R_W\frac{\sqrt{1+\tau i\omega}}{\sqrt{D_{\alpha}}(i\omega)^{1-\alpha/2}} ,
	\end{equation}
For $\tau=0$ and $\alpha=1$ Eq.~(\ref{imp3}) corresponds to the classical Warburg impedance \cite{mcdonald}.
For subdiffusive systems the relation on the impedance~(\ref{pulukl}) should be repleaced by  $\omega_{d}\equiv(D/L^{2})^{1/\alpha}$ \cite{bc}.
	
When $\omega\rightarrow\infty$, a substantial influence of $\tau$ can be inferred form Eqs.~(\ref{imp2}). 
When $\omega\rightarrow\infty$ from Eqs.~(\ref{imp2}) and (\ref{lambda}) we obtain
\begin{itemize}
		\item For $\tau\neq0$
	\begin{equation}\label{imp4}
Z(i\omega)=\frac{R_W\sqrt{\tau}}{\sqrt{D_{\alpha}}\omega^{(1-\alpha)/2}}\left[\cos\left(\pi\frac{1-\alpha}{4}\right)-i\sin\left(\pi\frac{1-\alpha}{4}\right)\right] ,
	\end{equation}
and the Nyquist plot of the impedance is a linear function passing through the origin of coordinates with the angle slope $\varphi$ given by
	\begin{equation}\label{tg1}
\varphi=\pi\frac{1-\alpha}{4} .
	\end{equation}
Let us note that for $0<\alpha<1$ there is $\varphi\in(0,\pi/4)$.
		\item For $\tau=0$
	\begin{equation}\label{imp5}
Z(i\omega)=\frac{R_W}{\sqrt{D_{\alpha}}\omega^{1-\alpha/2}}\left[\cos\left(\pi\frac{1-\alpha/2}{2}\right)-i\sin\left(\pi\frac{1-\alpha/2}{2}\right)\right] ,
	\end{equation}
where
	\begin{equation}\label{tg2}
\varphi=\pi\frac{1-\alpha/2}{2} ,
	\end{equation}
and $\varphi\in(\pi/4,\pi/2)$.
\end{itemize}

For low $\omega$ one obtains
	\begin{equation}\label{a}
Z(i\omega)=-\frac{R_W}{\lambda(i\omega)}\left[\frac{a_L\lambda(i\omega)-b_L L\gamma(i\omega)+a_L L\lambda(i\omega)\gamma(i\omega)}{a_L\lambda(i\omega)+b_L L\gamma(i\omega)-a_L L\lambda(i\omega)\gamma(i\omega)}\right] .
	\end{equation}
An analysis of Eq.~(\ref{a}) leads to the following conclusions. For $\omega\rightarrow\infty$ and $\alpha\in(0,1)$ that the slope of the plot is $\varphi=\pi(1-\alpha)/2$ when $b_L\neq0$ (for the partially or fully absorbing wall) and $\varphi=\pi/2$ when $b_L=0$ (for reflecting wall). We note that for $\omega\ll 1/\tau$ the terms which contain the $\tau$ parameter  can be neglected in the above formulas. Thus, in all cases $\varphi$ is also independent of $\tau$. 
  
\begin{figure}[h!]
	\centering
		\includegraphics{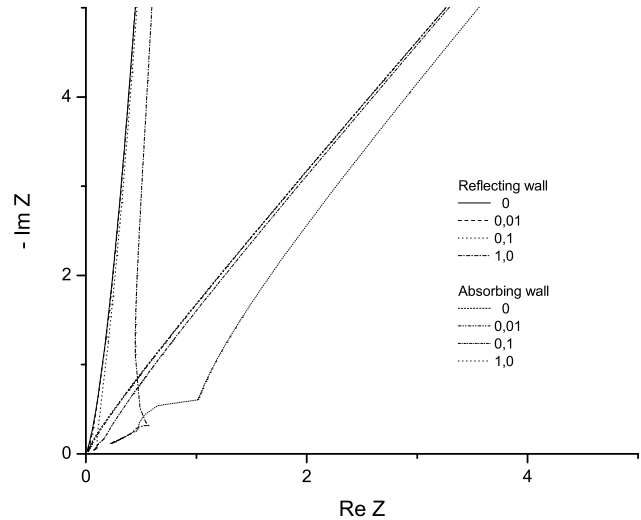}
	\caption{The Nyquist plots for $\alpha=0.4$ and for $\tau$ given in the legend.}
	\label{rys1}
\end{figure}

\begin{figure}[h!]
	\centering
		\includegraphics{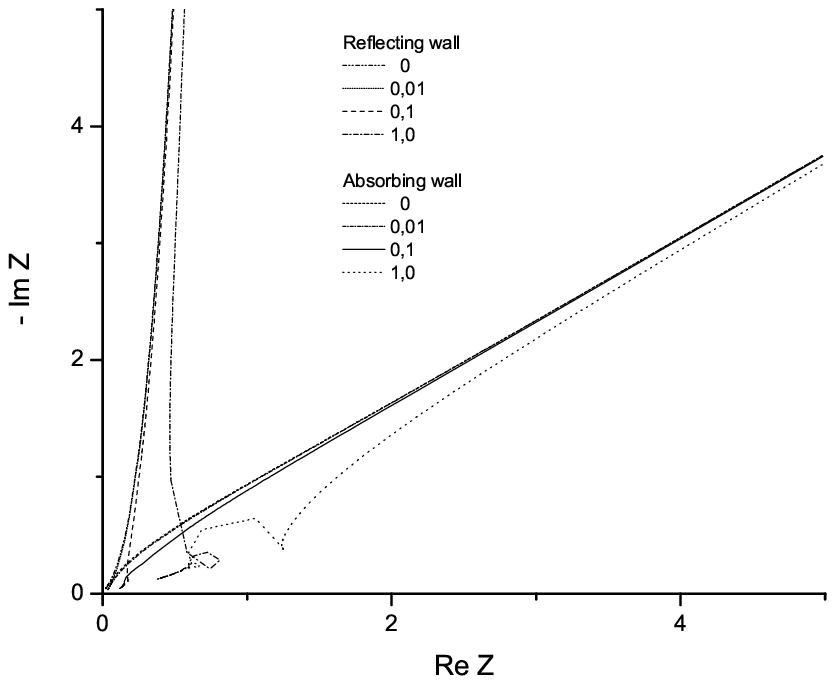}
	\caption{The Nyquist plots for $\alpha=0.6$.}
	\label{rys2}
\end{figure}

\begin{figure}[h!]
	\centering
		\includegraphics{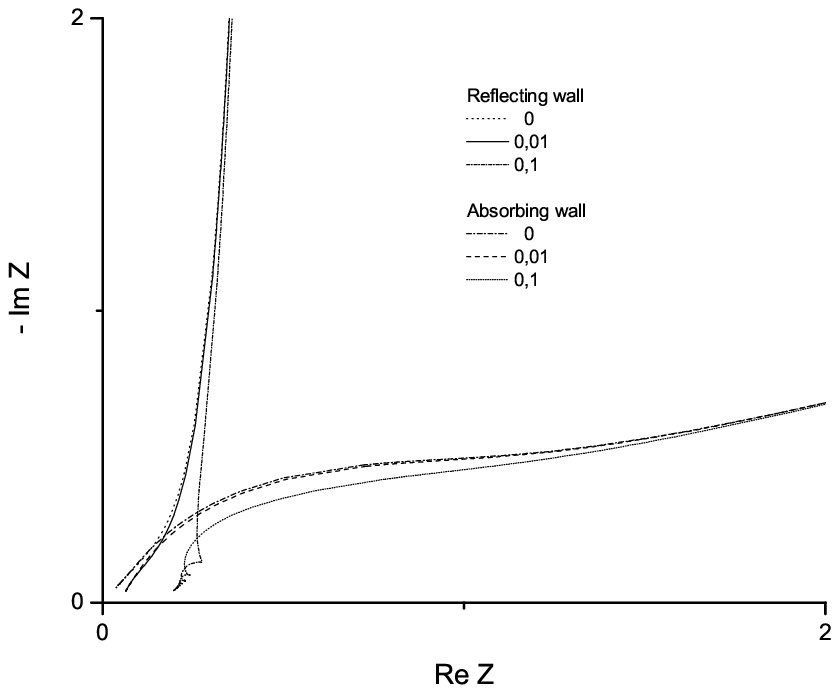}
	\caption{The Nyquist plots for $\alpha=0.8$.}
	\label{rys3}
\end{figure}

\begin{figure}[h!]
	\centering
		\includegraphics{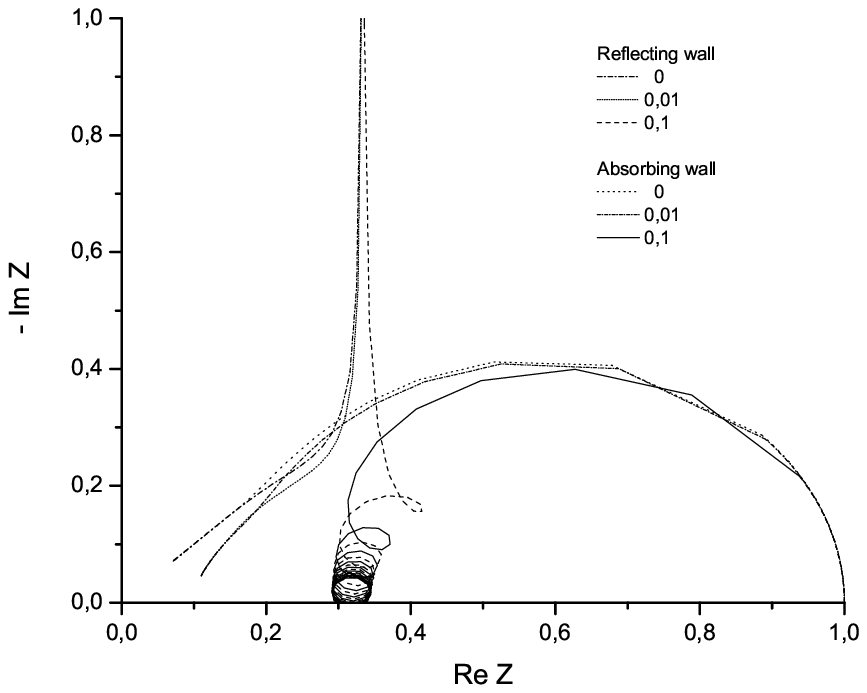}
	\caption{The Nyquist plots for the normal diffusion case ($\alpha=1.0$).}
	\label{rys4}
\end{figure}

Calculating $Re Z$ and $Im Z$ from (\ref{imp2}) for $s=i\omega$ we obtain the Nyquist plots (Figs.~\ref{rys1}--\ref{rys4}) with several values of the parameters $\tau$ and $\alpha$. Our calculations were done for $\omega\in(10^{-1},10^3)$, $R_W=1$, $L=1$ and $D_\alpha=1$ (all quantities are in arbitrary units). For larger values of $\omega$ the points on the plots are located near the origin and for larger $\tau$ the curves are located near the $Re Z$ axis.

On Figs. (\ref{rys1}) and (\ref{rys2}) the plots for $\tau=0$ and $\tau=0.01$ are practically undistinguishable, but for larger values of the subdiffusion parameter $\alpha$ these plots differ from each other. The curves with larger values of the $\tau$ parameter lie nearer the axis $Re Z$. The plots suggest that the curves for different values of $\tau$ converge into one curve for very small values of $\omega$. For relatively large values of $\tau$ the Nyquist plots show 'chaotic' behavior, which is stronger when $\alpha$ increases. Therefore, in the presented cases we did not consider values of $\tau$ larger than the one on plots in Figs.~(\ref{rys1}) -- (\ref{rys2}).

\section{Final remarks}

The main result of our work is Eq.~(\ref{imp2}) with the asymptotic formulas (\ref{imp4})--(\ref{a}). These functions illustrated by the plots Figs.~\ref{rys1}--\ref{rys4} show the following:

For $\omega\rightarrow\infty$ the Nyquist plot is the linear function passing through the origin of coordinates with the slop angle $\varphi$ dependent on subdiffusion parameter $\alpha$ and given by Eq.~(\ref{tg1}) and (\ref{tg2}). It is interesting that $\varphi$ does not depend on the parameter $\tau$ explicitly, but the dependence is `hidden' and is manifested by the interval to which the parameter belongs. For a given $\alpha\in(0,1)$ and $\tau\neq 0$ $\varphi=\pi(1-\alpha)/4<\pi/4$ and 
$\varphi=\pi(1-\alpha/2)/2>\pi/4$ when $\tau= 0$. 
For non-zero $\tau$, $\varphi$ does not dependent on $\tau$ and for all values of $\varphi$ it is independent of the boundary conditions at $x=L$. In the paper \cite{bisquert1998} the case of $\varphi\neq\pi/4$ was interpreted somewhat different, mainly as a presence of so-called `constant phase element' (CPE) in the system. 

For relatively high $\omega$ and for $\tau\neq 0$ the Nyquist plots show the `chaotic' and `oscillating' behavior, increases with increase of $\tau$ and $\alpha$ parameters. The physical interpretation of this fact can be as follows. Namely, the periodic changes of concentration at the surface $x=0$ generated the flux which is delayed in time by $\tau$ with respect to the concentration gradient. When the oscillations of concentration are very rapid, the flux does not keep up with concentration changes, so the additional factor contributing to the total impedance is created. The difficulties in movement of the ions increase when $\alpha$ decreases causing a decrease in the flux. Thus, the $\tau$ parameter  less influences the transport process when $\alpha$ is smaller. Such a behavior is observed on the presented plots, where the curves obtained for $\tau=0$ and for $\tau=0.1$ differ slightly form each other when $\alpha=0.4$ and $\alpha=0.6$, but the difference is relatively large for $\alpha=0.8$ and $\alpha=1$. 
For low $\omega$ the plots are dependent on the boundary conditions. According to the Eq. (\ref{a}), for fully reflecting wall and $\omega\ll 1/\tau$ one gets $\varphi=\pi/2$, which is independent of $\alpha$. For fully or partially permeable wall we observe that the plots become the linear where $\varphi=\pi(1-\alpha)/2$.

As we mentioned in the Introduction, there are a few methods to extract the value of subdiffusion parameters from experimental data. The considerations presented in this paper show that it is possible to determine the values of parameters of the system from the Nyquist plots obtained from experimental data. Measuring the slope angle for high $\omega$, one can extract the subdiffusion parameter $\alpha$ according to the formulas (\ref{tg1}) and (\ref{tg2}). Such a method was used to extract the parameter $\alpha$ for the lithium transport in gel electrode \cite{jpl}. The authors found that $\varphi=32^o$, it interpreted the results as subdiffusive transport of the particles inside the electrode. Since $\varphi<45^o$, we deduce that the transport studied in \cite{jpl} can be described by the Cattaneo equation with non-zero parameter $\tau$. Unfortunately, the value of slope angle is not sufficient to extract $\tau$ from experimental data. To find this parameter one should perform more detailed studies where the parameter representation of the Nyquist plots is taken into considerations.

For the system described by parabolic subdiffusion equation (\ref{pse}) one obtains $\varphi>\pi/4$. To achieve the situation where $\varphi<\pi/4$, the subdiffusion equation different from (\ref{pse}) was involved into the model \cite{bc}. Namely, it was assumed that the subdiffusion is described by one of the following equations
	\begin{equation}\label{e1}
\frac{\partial^\alpha C(x,t)}{\partial t^\alpha}=D_\alpha\frac{\partial^2 C(x,t)}{\partial x^2} ,
	\end{equation}
	\begin{equation}\label{e2}
\frac{\partial^{2-\alpha} C(x,t)}{\partial t^{2-\alpha}}=D_\alpha\frac{\partial^2 C(x,t)}{\partial x^2} ,
	\end{equation}
when both of the equations contain the Riemann-Liouville fractional derivative. However, physical meaning of Eqs. (\ref{e1}) and (\ref{e2}) is rather unknown since they were derived only on the phenomenological way, where the time derivative of natural order was replaced by the fractional one in the continuity equation and/or in the Fick's law. These equations were not derived on the base of `microscopic' model such as Continuous Time Random Walk formalism. We note that the Eq. (\ref{pse}) is equivalent to (\ref{e1}) if in the latter one the Riemann-Liouville fractional derivative is replaced by the Caputo derivative. In our paper we show that the slope $\varphi<\pi/4$ is achieved form the model based on hyperbolic subdiffusion equation, which has clear physical interpretation.

\section*{Acknowledgements}

This paper was supported by Polish Ministry of
Education and Science under Grant No. 1 P03B 136 30.

\end{document}